\documentclass[twocolumn]{aa}
\usepackage{graphicx}
\usepackage{txfonts}
\newcommand{\eo}{${\mathrm{E_0}}$}
\newcommand{\ep}{${\mathrm{E_{\mathrm p}}}$}
\newcommand{\erad}{E$_{\mathrm{rad}}$}
\newcommand{\al}{$\alpha$}
\newcommand{\be}{$\beta$}
\newcommand{\nga}{N$_{\mathrm \gamma}$}
\newcommand{\nsp}{N$_{\mathrm \gamma}$/E$_{\mathrm p}$}

\newcommand{\epo}{e$_{\mathrm p}$}
\newcommand{\ngao}{n$_{\mathrm \gamma}$}
\newcommand{\nspo}{n$_{\mathrm \gamma}$/e$_{\mathrm p}$}
\newcommand{\nspt}{n$_{\mathrm \gamma}$/e$_{\mathrm p}$/$\sqrt{\mathrm{t}_{90}}$}

\begin{document}
   \title{A simple empirical redshift indicator for gamma-ray bursts}

   \subtitle{}

   \author{J-L. Atteia
          \inst{1}
          }

   \offprints{J-L. Atteia}

   \institute{Laboratoire d'Astrophysique, Observatoire Midi-Pyr\'en\'ees, 31400 Toulouse, France 
              \email{atteia@ast.obs-mip.fr}             }

   \date{Received ; accepted }

   \abstract{We propose a new empirical redshift indicator for gamma-ray bursts.
This indicator is easily computed from the gamma-ray burst spectral parameters and
its duration, and it provides ``pseudo-redshifts'' accurate to a factor two.
Possible applications of this redshift indicator are briefly discussed.

   }

   \maketitle
%

\section{Introduction}
\label{intro}
Gamma-ray bursts (GRBs) are huge stellar explosions which have been 
observed at  redshifts ranging from 0.0085 to 4.5. 
While GRBs are in principle detectable out to very large redshifts ($z=10-20$, \cite{lamb00}), 
redshifts measured to date do not exceed 4.5.
The method most frequently used to measure GRB redshifts is to find a visible afterglow, 
and to identify absorption lines in its spectrum, caused by the gas in the GRB host galaxy. 
The redshift of the host can also be measured at late times 
from the host emission lines, when the afterglow has faded below detection.
Another, less frequent, method uses X-ray lines detected in 
the X-ray afterglows of some GRBs.
The absence of GRB detection beyond $z=5$ could be explained by the fact
that the afterglows of such distant GRBs must be searched for in the infrared,
due to the Lyman alpha cutoff.
The difficulty to measure spectroscopic redshifts led various authors to
propose alternate ways to determine GRB redshifts.
\cite{norr00} and \cite{reic01} have found empirical luminosity estimators 
based on GRB light curves. 
Such luminosity estimators can be used to infer the intrinsic luminosity
of individual GRBs, and consequently their redshifts.
While these estimators cannot be used to obtain precise redshifts for individual GRBs, 
they are useful to derive statistical properties of the GRB population.

Redshift estimators based on the gamma-ray data only present 
two distinctive advantages: they provide redshift estimates for most 
GRBs detected in gamma-rays, and they do not require extensive follow-up campaigns
involving large telescopes on the ground or in space.
Important issues can be addressed with moderatly accurate redshifts, like
the amount of energy released by GRBs in gamma-rays, the luminosity function
of GRBs, or the history of stellar formation at high redshifts.

We propose here a new method to obtain redshift indicators for GRBs from
gamma-ray observations. 
Our method is calibrated with 17 GRBs detected with BeppoSAX (\cite{boel97}) and HETE (\cite{rick01}).
In the following we call the redshifts inferred from our redshift indicator ``pseudo-redshifts''.
Pseudo-redshifts have the advantage of being very easily computed.
In addition to the possible applications already mentioned, 
pseudo-redshifts may become a useful tool to quickly identify high-redshift GRBs. 

\section{An empirical redshift indicator}
\label{redshift}

Finding redshift indicators for GRBs based on the gamma-ray data alone has always faced the
problem of the large intrinsic dispersion of GRB properties. 
This intrinsic dispersion prevents the determination of the redshifts
of individual GRBs.
With the measure of an increasing number of GRB redshifts it appeared, however,
that several properties of GRBs are correlated with the isotropic-equivalent
energy radiated in gamma-rays (called \erad\ in the following).
For instance, the correlation of the spectral hardness with \erad\ has been suspected for a long
time (see e.g.  \cite{atte00}, and \cite{lloy00}). 
It has only been demonstrated recently by Amati et al. (2002) for 12 GRBs with known redshifts.
The correlation of the duration with \erad\ is discussed in \cite{lee00}.
These correlations have led some authors to propose using the observed GRB properties to infer
\erad , and then to deduce the redshift from the comparison of the
observed fluence with \erad . 
\cite{norr00}, for instance, estimate \erad\ from the magnitude 
of the time lags between a high energy band and a low energy band. 
\cite{reic01} estimate \erad\ from the variability of the light curve.
We propose and test here another approach: we search a quantity which depends little 
on \erad , and which has a small intrinsic dispersion which does not blur the 
dependence on redshift.
Starting from empirical considerations, We find such a quantity essentially
based on the spectral characteristics of GRBs.

GRB energy spectra are well fit with the so-called GRB model, consisting of two
smoothly connected power laws (\cite{band93}).
In the following, $\alpha$ is the index of the low-energy power law, $\beta$ 
the index of the high-energy power law, and \eo\ is the break energy.
With this parametrization, the peak energy of the $\nu f_{\nu}$ spectrum
is $E_{\mathrm p} = E_0 \times (2+\alpha)$.
\ep\ is well defined for $\alpha \ge -2$ and $\beta < -2$.

Our method is based on the recent finding by Amati et al. (2002)
of a correlation between the intrinsic (redshift corrected) 
\ep\ of 12 GRBs with known redshifts, and \erad , their isotropic-equivalent energy 
radiated in gamma-rays.
According to Amati et al., \ep\ is roughly proportional to the square-root
of $E_{rad}$.
Since \al\ and \be\ do not vary too much from burst to burst, and since the 
energy radiated in gamma-rays is more or less the product of the number of photons
by their typical energy, we make the assumption that {the isotropic-equivalent 
number of photons in a GRB, \nga ,
is also roughly proportional to the square-root of \erad }.
For this study, we define \nga\ as the number of photons below the break,  
integrated from \ep /100 to \ep /2 .

\begin{figure}[t]
\resizebox{\hsize}{!}{\includegraphics{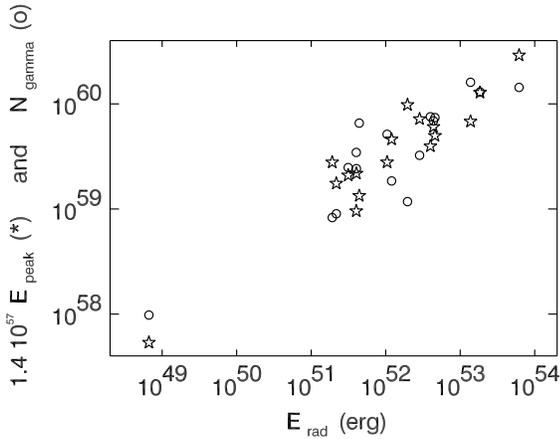}}
\caption{Dependence of two characteristic GRB quantities 
on \erad , the isotropic-equivalent energy radiated in gamma-rays.
The stars show the dependence of the intrinsic peak energy (multiplied here by 1.4 10$^{57}$). 
The circles show the dependence of the isotropic-equivalent number of photons emitted by the source.}
\label{fig1}
\end{figure}

\begin{figure}[t]
\resizebox{\hsize}{!}{\includegraphics{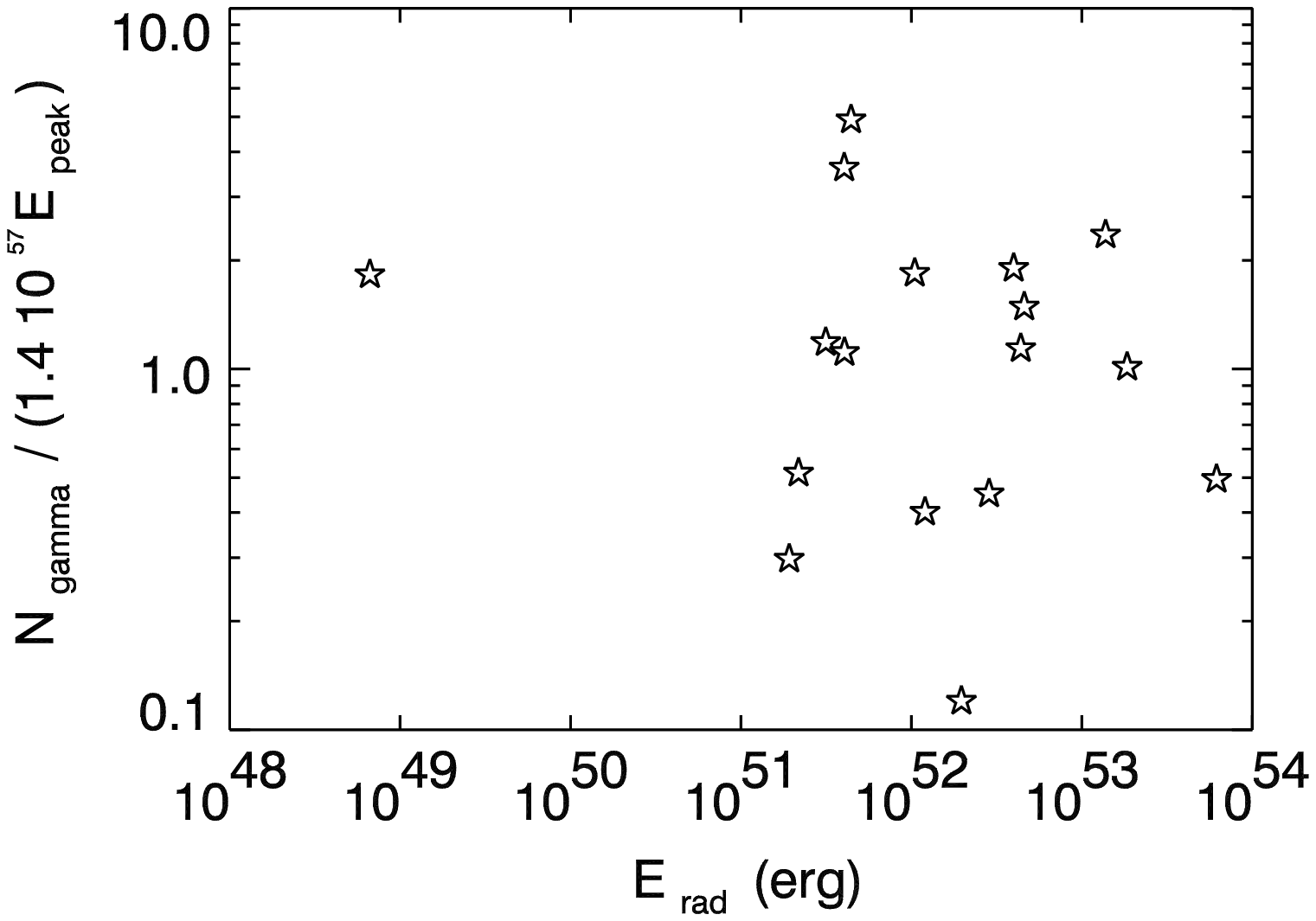}}
\caption{The ratio \nsp\ (see text) as a function of \erad .
This figure illustrates the weak dependence of \nsp\ with \erad.}
\label{fig1b}
\end{figure}

Fig. \ref{fig1} shows \ep , the intrinsic peak energy, and \nga , the 
isotropic-equivalent number of photons as a function of \erad\ for a sample of 17 GRBs detected
by BeppoSAX, BATSE, and HETE. 
The main characteristics of these GRBs are given in Table \ref{tab1}, along
with references for their spectral parameters.
The redshifts have been taken from J. Greiner's GRB page at
http://www.mpe.mpg.de/\~{}jcg/grbgen.html (except for GRB 020124, which comes from
\cite{hjor03}).
Figure \ref{fig1} shows that, as we suspected,  \ep\ and \nga\ have roughly the
same dependence on \erad .
We can thus go one step further with our main conjecture: we suppose 
that the ratio \nsp\ is almost independent of \erad , and can
be used as a redshift indicator.
Fig. \ref{fig1b} shows that indeed the ratio \nsp\ shows very little dependence on \erad, confirming
our conjecture.
This is not sufficient, however, to make it a correct redshift indicator. 
The critical issue is to find an indicator which has a small dependence on \erad ,
a strong dependence on the redshift, and a not too strong intrinsic dispersion. 
This issue is discussed in the next section.

\subsection{Definition of the redshift indicator}
\label{definition}

The theoretical considerations in the previous section are based on the study of 
{\it intrinsic} GRB properties.
Defining a redshift indicator implies that we do not know the redshifts of the GRBs
which are being studied, but only their observed properties.
To keep in mind this difference, in all the following we use capital letters for intrinsic 
quantities, and lower case for observed quantities.

As discussed earlier, the best redshift indicator is not necessarily
the one with the smallest intrinsic dispersion, but rather the one which has the 
best combination of a small intrinsic dispersion and a large dependence on redshift.
Relying on the analysis of the previous section, we propose to base our redshift indicator
on \nspo\, the ratio of the observed number of photons in the GRB on the 
observed peak energy. 
We tried various simple combinations of GRB parameters, all involving the ratio \nspo ,
and found that $X = $ \nspt\  has the right combination of properties for a redshift indicator.
In this equation \epo\ is the observed peak energy, \ngao\ the observed number of photons
between \epo /100 and \epo /2, and $\mathrm{t}_{90}$ the observed duration.
We do not claim here that $X$ is definitely the best redshift indicator, we nevertheless
believe that it is sufficiently good to deserve further discussion.

We derive pseudo-redshifts from the measure of $X$ in the following way:
in a first step we compute the theoretical evolution
of $X$ with redshift; then we invert this relation to derive 
a pseudo-redshift from the observed value of $X$.
The evolution of $X$ with redshift can be written as $$X = A \times f(z)$$
$A$ is a constant of normalization, and $f$ describes 
the evolution of $X$ with redshift for a ``standard'' GRB 
($\alpha = -1.0$, $\beta = -2.3$, and \eo = 250 keV) in a ``standard'' universe
(H$_0$ = 65 km s$^{-1}$ Mpc$^{-1}$, $\Omega_0 = 0.3$, $\Omega_\Lambda = 0.7$).
GRB spectral parameters are not critical here, because we have shown in the previous section
that the ratio \nsp\ does not vary much from burst to burst.
The normalization constant $A$ has been chosen to have about the
same number of GRBs below and above the theoretical curve in Fig. \ref{fig2} ($A=60$).

Pseudo-redshifts $\hat{z}$ are then defined by 

\begin{equation}
\hat{z} = {1 \over A} \times f^{-1}(X)
\end{equation}

Their use as redshift indicators is discussed below.

\begin{table}[t]
\caption{Observed properties of 17 GRBs with known redshift. 
The ten columns give the GRB name, the duration 
T$_{90}$ in seconds, the three spectral parameters ($\alpha , \beta$, and E$_0$), 
the gamma-ray fluence S$_{\mathrm \gamma}$ in units of 10$^{-6}$ erg cm$^{-2}$, 
the spectroscopic redshift $z$, the pseudo-redshift $\hat{z}$, the ratio $\hat{z}$/$z$, 
and a reference for the spectral parameters.}
\label{tab1}
$$
\begin{array}{llllllllll}
\hline
\noalign{\smallskip}
\mathrm{Name} & 
\mathrm{T}_{90} & 
{\alpha} & 
{\beta}^{\mathrm{a}} &
\mathrm{E}_0 & 
\mathrm{S}_{\gamma} & 
z & 
\hat{z} & 
\hat{z} / z &
\mathrm{ref.} \\
     & 
\mathrm{sec}  & 
     &
     &
\mathrm{keV}  & 
     &
\mathrm{}  & 
\mathrm{}  &
     &      
     \\
\noalign{\smallskip}
\hline 
\noalign{\smallskip}
 970228              &   80 & -1.54 &  -2.5 &  250 &   11 & 0.695 & 0.94 & 1.36 & 1 \\
 970508              &   20 & -1.71 &  -2.2 &  275 &  1.8 & 0.835 & 0.95 & 1.14 & 1 \\
 971214              &   35 & -0.76 &  -2.7 &  125 &  8.8 &  3.42 & 2.87 & 0.84 & 1 \\
 980613              &   20 & -1.43 &  -2.7 &  163 &  1.0 & 1.096 & 2.19 & 2.00 & 1 \\
 990123              &  100 & -0.89 & -2.45 &  703 &  300 &  1.60 & 2.20 & 1.38 & 1 \\
 990510              &   75 & -1.23 &  -2.7 &  210 &   19 &  1.619 & 1.44 & 0.89 & 1 \\
 990705              &   42 & -1.05 &  -2.2 &  199 &   75 &  0.843 & 0.85 & 1.01 & 1 \\
 990712              &   20 & -1.88 & -2.48 &  545 &  6.5 &   0.43 & 0.31 & 0.73 & 1 \\
 000131              & 96.3 &  -1.2 &  -2.4 &  163 &   26 &    4.5 & 1.35 & 0.30 & 2 \\
 010921              & 24.6 & -1.49 & -2.3 &  206 & 10.2 &   0.45 & 0.68 & 1.51 & 3 \\
 020124              & 78.6 & -1.10 &  -2.3 &  133 &  6.8 &    3.2 & 2.17 & 0.68 & 3 \\
 020813              &   90 & -1.05 &  -2.3 &  223 &  102 &   1.25 & 0.91 & 0.73 & 3 \\
 020903              &  9.8 &  -1.0 &  -2.3 &    3 & 0.09 &   0.25 & 0.26 & 1.03 & 4 \\
 021211              &  3.0 & -.896 &  -2.3 &   52 &  .96 &   1.01 & 0.76 & 0.75 & 5 \\
 030226              & 76.8 & -0.95 &  -2.3 &  103 &  6.4 &   1.98 & 2.35 & 1.19 & 6 \\
 030328              & 140. &  -1.0 &  -2.3 &  110 & 26.9 &   1.52 & 1.15 & 0.76 & 6 \\
 030329              & 22.8 & -1.03 &  -2.26 &  59 &  118 &  0.168 & 0.17 & 0.99 & 7 \\

\noalign{\smallskip}
\hline
\end{array}
$$
\begin{list}{}{}
\item[$^{\mathrm{a}}$] $\beta$ has been frozen to -2.3 for HETE GRBs 010921 to 030228.
\item[1] Amati et al. 2002. Fluence measured in the range 40-700 keV.
\item[2] Andersen et al. 2000. Fluence measured in the range 28-1800 keV.
\item[3] Barraud et al. 2003. Fluence measured in the range 30-400 keV.
\item[4] Sakamoto et al. 2003. Fluence measured in the range 7-30 keV.
\item[5] Crew et al. 2003. Fluence measured in the range 7-30 keV.
\item[6] Lamb et al. 2003. Fluence measured in the range 30-400 keV.
\item[7] Vanderspek et al. 2003. Fluence measured in the range 7-30 keV.
\end{list}
\end{table}

\subsection{Evaluation of pseudo-redshifts}
\label{evaluation}

\begin{figure}[t]
\resizebox{\hsize}{!}{\includegraphics{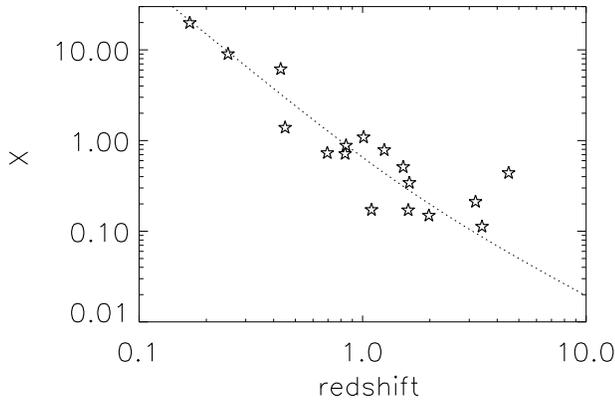}}
\caption{Correlation of $X$ = \nspt\ (see text ) with the measured redshifts of 17
 GRBs.
The isolated star at $z=4.5$ is GRB 000131.
The dotted line shows the theoretical relation between $\hat{z}$ and $z$.}
\label{fig2}
\end{figure}

Fig. \ref{fig2} shows the values of $X$
as a function of $z$, for the 17 GRBs of Table \ref{tab1}. 
This figure displays a clear anticorrelation between the two quantities.
The dotted line indicates the theoretical dependence $X = A \times f(z)$. 
The coefficient of correlation between $z$ and $X$ is $-0.875$, 
corresponding to a correlation significant
at the level of 4.9 sigmas using Fisher's Z transformation. 
We consider that this anticorrelation provides a good support to our intention
of using $X$ as a redshift indicator, 
and we use the equation (1) above to compute the pseudo-redshifts of GRBs in Table \ref{tab1}.
Table \ref{tab1} gives the values of $z$, $\hat{z}$, and their ratio, for the 17 GRBs 
with known redshift used in our analysis.\footnote{Two GRBs in \cite{amat02} are not included
in Table \ref{tab1} because they do not have spectroscopic redshifts. 
The redshift of GRB 980326 was estimated to be 1 from the observation of a supernova bump 
in the late light curve of the afterglow (\cite{bloo99}), and we find  $\hat{z}$=1.05. 
The redshift of GRB 000214 was estimated to be 0.42 from the observation of an iron line
in its X-ray afterglow (\cite{anto00}), and we find $\hat{z}$=0.39.}
It shows that $\hat{z}$ is usually within a factor of two of $z$, except for
GRB 000131 (at $z=4.5$), for which $z$ and $\hat{z}$ differ by a factor 3.3.
This discrepancy could be the consequence of the low quality of our redshift
indicator for this burst (most probably) or of a problem with the measure of the spectral parameters of this
GRB or of its redshift.
It might also indicate that the relation between $z$ and $\hat{z}$ is only
working (or calibrated) to $z=3.5$.
Because this event is clearly an outlier, we recomputed the coefficient
of correlation between $z$ and $X$ without it.
We find a coefficient of correlation of $-0.927$, 
corresponding to a correlation significant
at the level of 6.1 sigmas using Fisher's Z transformation.  

We conclude that the intrinsic dispersion of $X$ is not such that
it prevents its use as a redshift indicator.
We prefer the term redshift indicator than redshift estimator, because the ratio
of $\hat{z}$ over $z$ varies too much for a true redshift estimator. 
In the following we use the name pseudo-redshifts for $\hat{z}$.
Because $\hat{z}$ was derived from a purely empirical approach, we expect 
that an approach based on a physical treatment of GRB emission might provide
a better redshift estimator. 
While pseudo-redshifts are a potentially useful tool, they deserve a word
of caution here because we do not know how observational biases affect Figure \ref{fig2}.
We note for instance that GRBs with spectroscopic redshifts 
certainly represent a biased sample.
In addition the relations linking the GRB properties (from which we
compute $X$) to those of their afterglows (from which we measure $z$)
are far from being understood.
Anyone using this tool should thus keep in mind that Figure \ref{fig2} 
provides a biased view of the true distribution of GRBs in the ($z$,$X$) plane.

\section{An example of using pseudo-redshifts}
\label{application}

In this section we compute the pseudo-redshifts of 18 GRBs 
detected by HETE/FREGATE, whose spectral parameters are given in \cite{barr03}.
We compare them with the pseudo-redshifts of 8 GRBs with known redshifts 
in Table \ref{tab1} in order to assess the role of the redshift in the non-detection 
of the afterglows for these GRBs. 
The pseudo-redshifts of these 18 GRBs are given in Table \ref{tab2}.

The first remark is that short/hard GRBs should probably not be integrated in our framework.
GRB 020531 for instance has a low $X$ value, which results in an
unrealistically high pseudo-redshift. 
Having no redshift for short/hard bursts we cannot evaluate, and
eventually calibrate, our redshift indicator for these bursts.
The two shord/hard GRBs of our sample, GRB 020113 and GRB 020531, 
are thus excluded from the rest of our analysis.

\begin{table}[t]
\caption{Observed properties of 18 GRBs with no measured redshift. 
The eight columns give the name of the GRB, the time of arrival, the duration 
T$_{90}$ in seconds, the three spectral parameters ($\alpha , \beta$, and E$_0$), 
the gamma-ray fluence S$_{\gamma}$ in units of 10$^{-6}$ erg cm$^{-2}$ in the range 30-400 keV,
the pseudo-redshift $\hat{z}$, and a comment on the eventual detection of an afterglow
(XRR stands for X-Ray Rich GRB, OA, XA, and RA, respectively for Optical Afterglow, X-ray Afterglow,
and Radio Afterglow).}
\label{tab2}
$$
\begin{array}{llllllll}
\hline
\noalign{\smallskip}
\mathrm{Name} & 
\mathrm{Time} & 
\mathrm{T}_{90} & 
{\alpha} & 
\mathrm{E}_0 & 
\mathrm{S}_{\gamma} ^{\mathrm{e}} & 
\hat{z} &
\mathrm{Comment}  \\
     & 
\mathrm{SOD}  & 
\mathrm{sec}  & 
     &
\mathrm{keV}  &
     &
     &
     \\
\noalign{\smallskip}
\hline 
\noalign{\smallskip}
001225  & 25759 & 32.3 & -1.17 &  283 &  114 & 0.64 &  \\
010126  & 33162 &  7.7 & -1.06 &  115 &  3.0 &  1.6 &  \\
010326A & 11701 & 23.0 & -.894 &  260 &   16 &  2.8 &  \\
010613  & 27235 & 152. & -1.40 &  176 & 20.3 & 0.85 &  \\
010629  & 44468 & 15.1 & -1.17 &   59 &  2.6 & 0.76 &  \mathrm{XRR} \\
010928  & 60826 & 48.3 & -.623 &  260 &   21 &  4.9 &  \\
020113  &  7452 & 1.31 & -0.46 &  239 &  1.3 &  2.3 &  \mathrm{Short/Hard} \\
020127  & 75444 &  9.3 & -1.19 &  156 &  0.9 &  3.9 &  \mathrm{XA, RA, host} \\
020214  & 67778 & 27.4 & -.256 &  176 &   93 &  1.7 &  \\
020305  & 42925 & 250. & -.861 &  143 & 10.4 &  4.6 &  \mathrm{OA, host} \\
020331  & 59548 & 56.5 & -.922 &  120 &  4.5 &  3.4 &  \mathrm{OA} \\
020418  & 63789 & 7.54 & -1.10 &  240 & 13.9 &  1.3 &  \\
020531  &  1578 & 1.15 & -1.10 &  810 &  1.2 & 13.5 &  \mathrm{Short/Hard} \\
020801  & 46721 & 336. & -1.32 &  116 & 16.3 & 0.95 &  \\
020812  & 38503 & 27.5 & -1.03 &  125 &  2.3 &  3.4 &  \\
020819  & 53855 & 33.6 & -1.03 &   94 &  5.4 &  1.5 &  \mathrm{RA, host} \\
021016  & 37740 & 81.6 &  -.98 &  132 & 11.3 &  2.1 &  \\
021104  & 25262 & 19.7 &  -1.0 &   27 &  0.3 &  1.6 &  \mathrm{XRR} \\

\noalign{\smallskip}
\hline
\end{array}
$$
\end{table}

The median pseudo-redshift of long GRBs in Table \ref{tab2} is 1.65,
while it is only 0.88 for the 8 FREGATE GRBs with a measured redshift in Table \ref{tab1}.
If we believe the correlation between the pseudo-redshifts and the true redshifts,
this indicates that the redshift certainly plays a role in the non-detection
of the afterglows of FREGATE GRBs, even if this is not the only factor as emphasized
by \cite{crew03}.

While pseudo-redshifts can be useful for statistical analyses, the information
they convey is probably not meaningful for individual GRBs.
We believe however that pseudo-redshifts could become a useful tool to quickly
identify high redshift GRBs from the gamma-ray data alone.
A first step in this direction is obviously to prove the validity of pseudo-redshifts
for this task.
GRB 020127 may appear as a good test case in this context because it has
a high pseudo-redshift ($\hat{z} = 3.9$ in Table \ref{tab2}), a possible X-ray afterglow, 
a possible radio afterglow, and a candidate host galaxy (\cite{fox02a}, 2002b, 2002c).
If the host candidate is at a redshift of about 3, this would strengthen the validity
of pseudo-redshifts as a tool for the quick identification of high redshift GRBs.

\section{Conclusion}
\label{conclusion}

We propose an empirical redshift indicator for GRBs, which is easily computed
from the gamma-ray data alone and provides ``pseudo-redshifts'' accurate to a factor of two.
Despite their moderate accuracy, we believe that their easy computation 
will make these pseudo-redshifts useful in future GRB studies. 
Their possible applications include a statistical comparison
of the distance distribution of distinct GRB populations, constraints on
the star formation rate at high redshifts, and the fast identification of 
remote GRBs, with redshifts beyond three.  

The usefulness of these pseudo-redshifts will ultimately depend on the 
confirmation of their accuracy, which will be tested as a larger number 
of GRBs with known redshifts become available.

\begin{acknowledgements}
The author is grateful to R. Mochkovitch for insightful discussions
on pseudo-redshifts and to C. Barraud for helpful suggestions.
JLA acknowledges the use of J. Greiner GRB page at http://www.mpe.mpg.de/\~{}jcg/grbgen.html,
and useful comments from the referees.
\end{acknowledgements}

\end{document}